\documentclass[12pt]{article}
\usepackage{epsfig,a4}
 \advance\oddsidemargin by -1in
 \advance\evensidemargin
 by -1in
 \marginparsep
 0.4cm
 \advance\topmargin by
 -0.0in
 \makeindex

 \newcommand\ra{\rangle}
 \newcommand\beq{\begin{equation}}
 
 \newcommand\eeq{\end{equation}}                                               
 \newcommand\beqn{\begin{eqnarray}}
 \newcommand\eeqn{\end{eqnarray}}

\def\BA{\begin{eqnarray}}
\def\BE{\begin{equation}}
\def\BF{\begin{figure}[htb]}
\def\BT{\begin{table}[htb]}
\def\EA{\end{eqnarray}}
\def\EE{\end{equation}}
\def\EF{\end{figure}}
\def\ET{\end{table}}

\def\ra{\rangle}

\def\lsim{\mathrel{\rlap{\lower4pt\hbox{\hskip1pt$\sim$}}
    \raise1pt\hbox{$<$}}}         
\def\gsim{\mathrel{\rlap{\lower4pt\hbox{\hskip1pt$\sim$}}
    \raise1pt\hbox{$>$}}}         

\begin{document} 
\vspace*{3cm}
 
\date{today} 

\begin{center}
{\Large\bf Dipole description of inclusive particle production}

\end{center}

 
\begin{center}
 
\vspace{0.5cm}
 {\large B.Z.~Kopeliovich$^{1,2}$, 
Ivan~Schmidt$^{1}$, A.V.~Tarasov$^{2}$,
O.O.~Voskresenskaya$^{2}$}
 \\[1cm]
 $^{1}${\sl Departamento de F\'\i sica, Universidad  
T\'ecnica Federico Santa Mar\'\i a,\\
Casilla 110-V, Valpara\'\i so, Chile}

 $^{2}${\sl Joint Institute for Nuclear Research, Dubna, 141980 Moscow
Region, Russia}\\[0.2cm]
 
\end{center}

\begin{abstract}

The effects of multiple interactions in colliding particles (e.g.  
in nucleus-nucleus collisions) are modeled using the light-cone dipole
approach. Guided by the abelian analogue of multi-photon interactions in
the production of a pair of charged particles, we relate the inclusive
cross section of quark pair production with the cross sections of
interaction of a QCD dipole with either the beam or the target.

\end{abstract}
 
  
\section{Light-cone dipole representation}

The light cone dipole description of hadronic interactions \cite{zkl,al}
offers quite an effective phenomenology. The central quantity of this
approach, the universal and flavor independent cross section of
interaction of a colorless dipole (quark-antiquark, or glue-glue) with a
target (proton) is fitted to data, and therefore it incorporates
information on all possible gluonic exchanges and bremsstrahlung including
also nonperturbative effects. This may be considered as an alternative to
the parton distribution function, with the advantage that it includes by
default all higher order corrections and higher twist effects. This
approach is especially powerful for calculating nuclear effects and
diffractive processes \cite{zkl,krt1,krt2,kps1}. Since QCD dipoles are
eigenstates of interaction, multiple interactions effects for the elastic
amplitude can be included via simple eikonalization.

The main difficulty of the dipole approach, unresolved so far, is modeling
the distribution amplitude of QCD dipoles in the projectile high energy
particle. This problem has only been solved in the lowest order of
perturbative QCD for a photon projectile \cite{bjorken,nz}, and for
radiation of photons and gluons by a color charge \cite{hir,km,kst1}.
However, once the multi-gluon exchange interactions with the target are
important, one should not restrict oneself to a single parton density in
the projectile, as is frequently done. Inclusion of higher order
corrections and soft multiple interactions in the projectile particle
remains a challenge.  A model for simultaneous inclusion of multiple
interaction effects both in the beam and target was constructed in
\cite{yuri} for gluon radiation in nucleus-nucleus collision. Here we
present another attempt to make progress in this direction.

The paper is organized as follows. In Sect.~\ref{analogue} we study the
production of a pair of charge particles employing abelian dynamics. We
start with the Born approximation (Sect.~\ref{born-qed}) and then include
multi-photon exchanges with the target, and eventually with both colliding
particles $A$ and $B$ (Sect.~\ref{multi-photon}).

In Sect.~\ref{quarks} we develop the formalism for nonabelian dynamics.
The Born approximation is described in Sect.~\ref{born-qcd}, and the
results are generalized to include multiple interactions in
Sect.~\ref{multi-gluon}.

\section{QED analogue}\label{analogue}

\subsection{Born approximation in QED}\label{born-qed}

\begin{figure}[tbh]
 \includegraphics{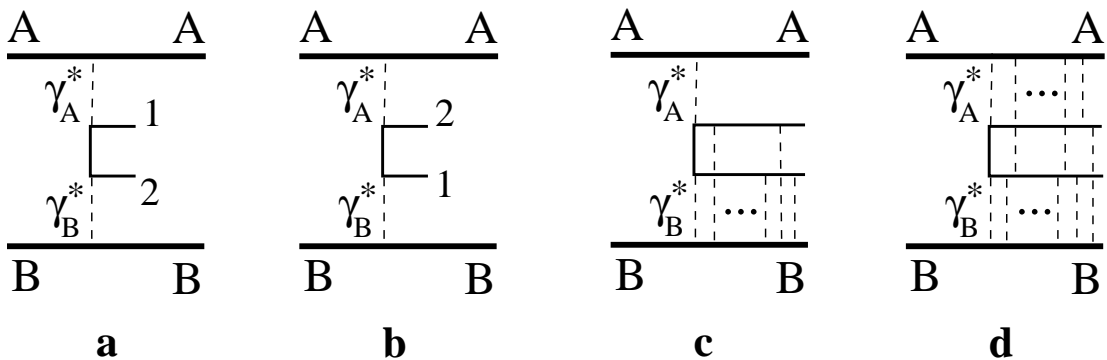}
 \begin{center} \vspace{5.5cm}
 \parbox{14cm} {\caption[Delta] {Production of charged particles $1$ and
$2$, in the Born approximation ($\bf a$ and $\bf b$), and including
multi-photon exchanges with the target or with both colliding particles
($\bf c$ and $\bf d$ respectively).}
 \label{qed}} 
\end{center}
 \end{figure}

We start with an abelian analogue for quark pair production, since this
process has a simpler dynamics, but it contains many features of the
nonabelian description. The process under discussion is production of a
pair of particles, $1$ and $2$, in the collisions of two hadrons (or
nuclei), $A$ and $B$,
 \beq
A +B \to  A +B + 1 + 2\ .
\label{1.10}
 \eeq
 For the sake of simplicity we assume that the produced particles are
spinless. The cross section of this process reads,
 \beq
d\sigma_{AB} = |M|^2\,d\Gamma \,
\label{1.20}
 \eeq
 where $M$ is the Lorentz-invariant amplitude, and $d\Gamma$ is a phase 
space element. In the lowest order of perturbative expansion corresponding 
to the graphs shown in Fig.~\ref{qed}a,b 
 the amplitude has the form,
 \beq
M=J_\mu^{(A)}(p_A,q_A)\,J_\mu^{(B)}(p_B,q_B)\,
\frac{t_{\mu\nu}(q_A,q_B,p_1,p_2)}{Q_A^2Q_B^2}\ ,
\label{1.30}
 \eeq
 where
 \beqn
q_{A(B)} &=& p_{A(B)}-p^\prime_{A(B)}\ ;\nonumber\\
Q^2_{A(B)} &=& -q_{A(B)}^2 \approx
\vec q^{\,2}_{A(B)_\perp} + q^2_{A(B)_{min}}\ ;\nonumber\\
q^2_{A(B)_{min}} &=& \frac{q_{A(B)}^2}{\gamma_{A(B)}^2}\ ;
\label{1.40}
 \eeqn
 $p_{A(B)}$, $p_{A(B)}^\prime$, $p_1$ and $p_2$ are the 4-momenta of
particles $A(B)$ in the initial and final states, and of the produced
particles respectively; $\gamma_{A(B)}$ are the Lorentz factors of the
colliding particles; $q_{A(B)}$ are the 4-momenta of the virtual photons
emitted by the particles $A(B)$. The electromagnetic currents in
Eq.~(\ref{1.30}) have the form,
 \beqn
&&J^{(A)}_\mu(p_A,q_A) =
\sqrt{4\pi\alpha_{em}}\,Z_A
F^{(A)}(Q_A^2)(2p_A-q_A)_\mu\ ;
\label{1.50}\\
&&J^{(A)}_\nu(p_B,q_B) =
\sqrt{4\pi\alpha_{em}}\,Z_B
F^{(B)}(Q_B^2)(2p_B-q_B)_\nu\ ,
\label{1.60}
 \eeqn
 where $Z_{A(B)}$ are the charges of $A$ and $B$, and
$F^{A(B)}(Q_{A(B)}^2)$ are their form factors, respectively.

The tensor $t_{\mu\nu}$ in (\ref{1.30}) reads,
\beqn
&&t_{\mu\nu} = 4\pi\alpha_{em}\left[
D_1^{-1}\,(2p_1-q_A)_\mu(2p_2-q_B)_\nu +
D_2^{-1}\,(2p_2-q_A)_\mu(2p_1-q_B)_\nu -
2g_{\mu\nu}\right]\ ;
\label{1.70}\\
&&D_1 = m_2-(p_1-q_A)^2=m^2-(p_2-q_B)^2\ ;
\label{1.80}\\
&&D_2 = m_2-(p_2-q_A)^2=m^2-(p_1-q_B)^2\ ;
\label{1.90}
 \eeqn
 Notice that,
 \beq
(q_{A})_\mu\,t_{\mu\nu} = 
t_{\mu\nu}\,(q_{B})_\nu = 0\ .
\label{1.100}
 \eeq

The phase space factor in Eq.~(\ref{1.20}) is
 \beqn
d\Gamma &=& 
\frac{d^3p^\prime_A\,d^3p^\prime_B\,d^3p_1\,d^3p_2}
{64\,I\,(2\pi)^8\,E_A^\prime E_B^\prime\epsilon_1\epsilon_2}
\ \delta^{(4)}(p_A+p_B-p^\prime_A-p^\prime_B
-p_1-p_2)\nonumber\\ &=&
\frac{d^4q_A\,d^4q_B\,d^3p_1\,d^3p_2}
{64\,I\,(2\pi)^8\,\epsilon_1\epsilon_2}\ 
\delta^{(4)}(q_A+q_B-p_1-p_2)\,
\delta(2p_Aq_A+Q_A^2)\,
\delta(2p_Bq_B+Q_B^2),
\label{1.110}
\eeqn
 where
\beqn
I = \sqrt{(p_Ap_B)^2-M_A^2M_B^2}\ .
\label{1.120}
 \eeqn
 Here $M_{A(B)}$ and $E_{A(B)}^\prime$ are the colliding hadrons or nuclei
masses and energies; $m$ and $\epsilon_{1(2)}$ are the masses and energies
of the produced particles.

To simulate QCD, we introduce a charge screening effect in what follows, 
i.e.
consider the colliding particles as neutral dipoles with the screening
potential,
 \beq
V(r) = \frac{Z\alpha_{em}}{r}\,
e^{-\lambda r}\ .
\label{1.130}
 \eeq
 This can be also simulated by an effective photon mass $\lambda$,
replacing in (\ref{1.30}), $Q_{A(B)}^2\Rightarrow Q_{A(B)}^2+\lambda^2$.

Now, let us consider the collision of ultra-relativistic particles
($\gamma_{A(B)}\gg1$) in the c.m. frame. Neglecting corrections of order
$O(\gamma_{A(B)}^{-2})$ we have from (\ref{1.20}) - (\ref{1.120}),
 \beqn
\sigma &=& \int dq_A^+dq_A^-dq_B^+dq_B^-\,
\frac{d^3p_1\,d^3p_2}{\epsilon_1\,\epsilon_2}\,
\sigma_0^{A}(q_A)\sigma_0^{B}(q_B)\,
|U(q_A,q_B,p_1,p_2)|^2
\nonumber\\&\times&
\delta(q_A^-)\,\delta(q_B^+)\,
\delta(q_A^+ + q_B^+ - p_1^+ - p_2^+)\,
\delta(q_A^- + q_B^- - p_1^- - p_2^-)\ ,
\label{1.140}
 \eeqn
 where $p^{\pm}_{1(2)} = (p_{1(2)})_0 \pm (p_{1(2)})_z$ are the light-cone
momenta of particles $1(2)$ (the $z$-axis is chosen along the momenta of
$A,\ B$);  $\sigma_0^{A}(q_A)$ and $\sigma_0^{B}(q_B)$ are the
differential cross sections of elastic scattering of particles $1,\ 2$ on
hadrons (nuclei) $A$ or $B$ respectively,
 \beqn
&& \sigma_0^{A}(q_A) \equiv 
\frac{d^2\sigma[1(2)+A\to 1(2)+A]}
{d^2q_{A_\perp}} =
\left[\frac{2Z_A\alpha_{em}\,F^{A}(Q_A^2)}
{\vec q^{\,2}_{A_\perp} + \lambda^2}\right]^2\ ;
\label{1.160}\\
&& \sigma_0^{B}(q_B) \equiv
\frac{d^2\sigma[1(2)+B\to 1(2)+B]}
{d^2q_{B_\perp}} =
\left[\frac{2Z_B\alpha_{em}\,F^{B}(Q_B^2)}
{\vec q^{\,2}_{B_\perp} + \lambda^2}\right]^2\ .
\label{1.170}
 \eeqn
 Then, we can represent
 \beqn
&& \frac{d^3p_{1(2)}}{\epsilon_{1(2)}} =
d^2p_{1(2)_\perp}\,\frac{dp^+_{1(2)}}{p^+_{1(2)}} =
d^2p_{1(2)_\perp}\,\frac{dp^-_{1(2)}}{p^-_{1(2)}}\ ;
\label{1.180}\\
&& p^+_{1(2)}p^-_{1(2)} = 
m^2+\vec p^{\,2}_{1(2)_\perp}\ .
\label{1.190}
 \eeqn

It is convenient to introduce the fractions of light-cone momenta of the 
colliding virtual photons carried by the produced particle $1$,
 \beqn
\alpha_A &=& \frac{p_1^+}{q_A^+}\ ;\nonumber\\
\alpha_B &=& \frac{p_1^-}{q_A^-}\ ,
\label{1.200}
 \eeqn
 which are connected by the relation,
 \beq 
\alpha_{A(B)} = \frac{(1-\alpha_{B(A)})(m_1)^2_{\perp}}
{(1-\alpha_{B(A)})(m_1)^2_{\perp} + \alpha_{B(A)})(m_2)^2_{\perp}}
\label{1.210}
 \eeq

The amplitude $U(q_A,q_B,p_1,p_2)$ in (\ref{1.140}),
 \beq
U(q_A,q_B,p_1,p_2) = \left[\frac{p_1^-p_2^+}{D_1} +
\frac{p_2^-p_1^+}{D_2} - 1\right]\ ,
\label{1.220}
 \eeq
 is function of three 4-momenta (since $q_A+q_B=p_1+p_2$). Therefore, we
can choose as independent variables the three transverse momenta and one
of the light-cone fractions $\alpha_A$ (or $\alpha_B$). Selecting $\vec
q_{A_\perp}$, $\vec q_{B_\perp}$, $\vec p_{1_\perp}$ and $\alpha_A$, we
can represent $U$ as,
 \beq
U(\vec q_{A_\perp},\vec q_{B_\perp},\vec 
p_{1_\perp},\alpha_A) =
\Phi(\vec p_{1_\perp} - \alpha_A\vec q_{A_\perp};\vec 
q_{A_\perp},\alpha_A) -
\Phi(\vec p_{1_\perp} - \alpha_A\vec q_{A_\perp} - \vec q_{B_\perp};
\vec q_{A_\perp},\alpha_A)\ ,
\label{1.230}
 \eeq
 where
 \beqn
\Phi(\vec p_\perp;\vec q_\perp,\alpha) &=&
\Phi^T(\vec p_\perp;\vec q_\perp,\alpha) + 
\Phi^L(\vec p_\perp;\vec q_\perp,\alpha)\ ;
\label{1.240}\\
\Phi^T(\vec p_\perp;\vec q_\perp,\alpha) &=&
\frac{2\alpha(1-\alpha)\,\vec p_\perp\cdot\vec q_\perp}
{\vec p_\perp^{\,2} +\epsilon^2(\vec q_\perp,\alpha)}\ ,
\label{1.250}\\
\Phi^L(\vec p_\perp;\vec q_\perp,\alpha) &=&
\frac{\alpha(1-\alpha)(1-2\alpha)\,\vec q_\perp^{\,2}}
{\vec p_\perp^{\,2} +\epsilon^2(\vec q_\perp,\alpha)}\ ,
\label{1.260}\\
\epsilon^2(\vec q_\perp,\alpha) &=&
m^2 + \alpha(1-\alpha)\vec q_\perp^{\,2}\ .
\label{1.270}
 \eeqn

Thus, the cross section of pair production can be 
represented as,
 \beq
\sigma(A+B\to A+B+1+2) =
\int dq_A^+\,d^2q_{A_\perp}\,n_A(q_A)\,
\sigma(\gamma*_A+B\to 1+2+B)\ ,
\label{1.280}
 \eeq
 where 
 \beq
n_A(q_A) = \frac{\sigma_0^A(q_A)\,\vec q_{A_\perp}^{\,2}}
{(2\pi)^2\,\alpha_{em}\,q_A^+} =
\left(\frac{Z_AF^A}{Q_A^2+\lambda^2}\right)^2\,
\frac{\alpha_{em}}{q_A^+}\ ,
\label{1.290}
 \eeq
 is the density of equivalent photons \cite{weiz,will} in the projectile
$A$.

The virtual photoproduction cross section $\sigma(\gamma^*_A+B\to 
1+2+B)$ in (\ref{1.280}) can be expressed in terms of the dipole 
formalism as,
 \beq
\sigma(\gamma^*_A+B\to 1+2+B) = 
\int d^2r\,d\alpha_A\,
\left[|\Psi^T_A(\vec r,\alpha_A)|^2 +
|\Psi^L_A(\vec r,\alpha_A)|^2\right]\,
\sigma_B(r)\ .
\label{1.300}
 \eeq
 Here $\Psi^{T,L}$ are the light-cone wave functions of transversely or
longitudinally polarized photons with 4-momentum $q_A$,
 \beq
\Psi^{T,L}_A(\vec r,\alpha_A) = 
\sqrt{\frac{\alpha_{em}}{\alpha_A(1-\alpha_A)}}\,
\frac{1}{|\vec q_{A_\perp}|(2\pi)^2} \int d^2p_T\,
\Phi^{T,L}(\vec p_\perp;\vec q_{A_\perp},\alpha_A)\,
e^{i\vec p_\perp\cdot\vec r}\ .
\label{1.310}
 \eeq

The cross section of interaction of the dipole of particles $1-2$ with 
$B$ has the standard form \cite{zkl},
 \beq
\sigma_B(\vec r) = 2\int d^2q_{B_\perp}\,
\sigma_0^B(q_B)\,\left(1 -
e^{i\vec q_{B_\perp}\cdot\vec r}\right)\ .
\label{1.320}
 \eeq

The resulting representation, Eq.~(\ref{1.300}), which treats the
production of particles $1-2$ as photoproduction by a virtual photon
representing the electromagnetic field of the projectile $A$ interacting
with the target $B$, looks asymmetric relative to the replacement
$A\Leftrightarrow B$. This is, however, an artifact of our choice of
$\alpha_A$ as a variable. If our choice were $\alpha_B$, the same cross
section would look as a result of interaction of a photon $\gamma^*_B$
with the target $A$. Thus, the choice of an independent variable,
$\alpha_A$ or $\alpha_B$, leads to a breaking of the symmetry,
$A\Leftrightarrow B$.

\subsection{Multi-photon exchanges}\label{multi-photon}

So far our considerations were restricted to the Born (one photon)
approximation. It turns out, however, that the relation Eq.~(\ref{1.300})
is also correct if the dipole $1-2$ interacts with the target $B$ via
multiple-photon exchanges as is illustrated in Fig.~\ref{qed}c. This is
particularly important if $B$ is a nucleus.  Indeed, in this case the
dipole cross section reads,
 \beq
\sigma_B(r) = 2\int d^2b\,
\Bigl\{1 - \exp[i\Delta\chi_B(\vec b, \vec r)]\Bigr\}\ ,
\label{1.330}
 \eeq
 where the phase shift
 \beqn
\Delta\chi_B(\vec b, \vec r) &=&
\chi_B(\vec b_+)-\chi_B(\vec b_-)\ ;
\label{1.340}\\
\chi_B(\vec b_\pm) &=& \frac{Z_B\alpha_{em}}{\pi}
\int \frac{d^2q_\perp}{\vec q_\perp^{\,2} + \lambda^2}\ 
e^{i\vec q_\perp\cdot\vec b_\pm}\ ;
\label{1.350}\\
\vec b_+ &=& \vec b +(1-\alpha_A)\vec r\ ,
\nonumber\\
\vec b_- &=& \vec b -\alpha_A\vec r\ .\nonumber
 \eeqn
 At first glance the cross section Eq.~(\ref{1.330}) depends also on
$\alpha_A$. However, the change of integration variable, $\vec b
\Rightarrow \vec b +(\alpha_A-1/2)\vec r$ eliminates the $\alpha_A$
dependence.

Another possible representation for the dipole cross section has the form,
 \beq
\sigma_B(r) = 2\int d^2q_\perp\, 
\sigma_{Gl}^B(\vec q_\perp)\,
\left(1 - e^{i\vec q_\perp\cdot\vec r}\right)\ ,
\label{1.360}
 \eeq
 where $\sigma_{Gl}^B(\vec q_\perp)$ is the differential cross section of 
elastic scattering of one of the particles, $1$ or $2$, with the target 
$B$,
calculated in the eikonal (Glauber) approximation,
 \beqn
\sigma_{Gl}^B(\vec q_\perp) &=& 
\left|f_B^\pm(\vec q_\perp)\right|^2\;
\label{1.370}\\
f_B^\pm(\vec q_\perp) &=& \frac{i}{2\pi}
\int d^2b\,e^{i\vec q_\perp\cdot\vec b}\,
\left[1 - e^{\pm i\chi_B(\vec b)}\right]\ .
\label{1.380}
 \eeqn
 The signs $\pm$ correspond to the opposite charges of particles $1,\ 2$.

Thus, we conclude that the effect of all multi-photon exchanges with the
target $B$ (restricted to only one photon exchange with the projectile
$A$) is equivalent to the replacement of the Born cross section
$\sigma_0^B(q_B)$ by the Glauber one, $\sigma_{Gl}^B(q_B)$.

The same cross section of pair production calculated in a single photon
approximation for $A$, but multi-photon with $B$, can be
represented differently if one chooses $\alpha_B$ as a variable,
 \beq
\sigma(\gamma*_A+B\to 1+2+B) =
\int dq_{B_-}\,d^2q_{B_\perp}\,
\tilde n_B(q_B)\,\sigma(\gamma^*_B+A\to1+2+A)\ .
\label{1.390}
 \eeq
 Here the process $\gamma^*_B+A\to 1+2+A$ is calculated in one-photon 
approximation, but the density function of "equivalent photons"
is different from (\ref{1.290}),
 \beq
\tilde n_B(q_B) = 
\frac{\sigma_{Gl}^B(q_B)\,\vec q_{B_\perp}^{\,2}}
{(2\pi)^2\,\alpha_{em}\,q_B^-}\ ,
\label{1.400}
 \eeq
 with the replacement of single- to multi-photon exchange, 
$\sigma_{0}^B(q_B)\Rightarrow \sigma_{Gl}^B(q_B)$.

It is natural to assume that the inclusion of multi-photon exchanges
between the produced quark pair and both colliding nuclei
(Fig.~\ref{qed}d) can be done by replacing the "single-photon" quantity
$\sigma(\gamma^*_B + Z_A \to 1+2+Z_A)$ in (\ref{1.390}) by a multi-photon
one. To do that we should replace the Born approximation for the cross
section $\sigma_A(\vec r)$ in (\ref{1.300}) (with the interchange
$A\Leftrightarrow B$),
 \beq
\sigma_A(\vec r) = 2\int d^2q_A\,
\sigma^A_0(q_A)\,
\left(1-e^{i\vec q_A\cdot\vec r}\right)
\label{1.410}
 \eeq
 by the following Glauber form,
 \beq
\sigma_A(\vec r) = 2\int d^2b\,
\left[1 - e^{i\Delta\chi_A(\vec b,\vec r)}\right]\ .
\label{1.420}
 \eeq
 The phase shifts $\Delta\chi_A(\vec b,\vec r)$ are defined on analogy to 
$\Delta\chi_B(\vec b,\vec r)$ in (\ref{1.340})-(\ref{1.350}).

\section{Quark production}\label{quarks}

\subsection{Born approximation in QCD}\label{born-qcd}

\begin{figure}[tbh] 
\includegraphics{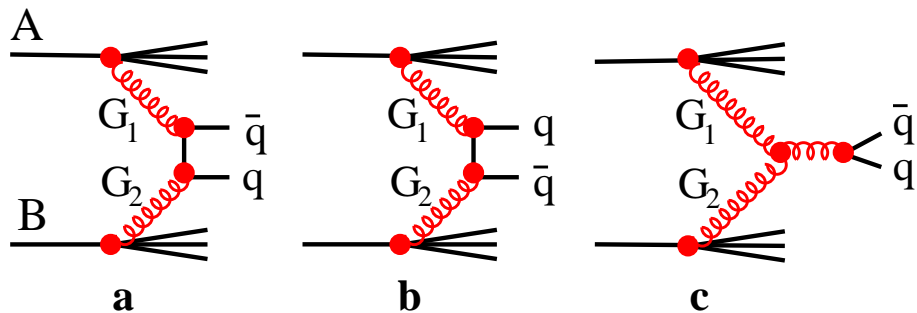} 
\begin{center} 
\vspace{4.5cm}
\parbox{14cm} {\caption[Delta] {One gluon approximation to the central 
production of a $\bar qq$ pair.}
 \label{one-gluon}} 
\end{center}
 \end{figure}

The Feynman graphs corresponding to $\bar qq$ pair production in a
collisions of two hadrons $A$ (beam) and $B$ (target) in lowest order in
$\alpha_s$ (double one-gluon approximation) are shown in
Fig.~\ref{one-gluon}.
 We assume that only one quark (antiquark) is detected, with transverse
momentum $\vec p_T$ and rapidity $y$, while the accompanying antiquark
(quark) is not observed, i.e. its momentum is integrated out. Then the
cross section corresponding to the graphs in Fig.~\ref{one-gluon} has the
form,
 \beq
\frac{d\sigma(A\,B\to q\,X)}{d^2p_T\,dy} = 
\frac{4\pi\alpha_s}{3}
\int\limits_{x_q}^1 \frac{dx_1}{x_1}\,
\int d^2q_1\,d^2q_2\,
\frac{q_1^2\,{\cal F}_A(x_1,\vec q_1)}
{(q_1^2+q_{1\,min}^2)^2}\,
\frac{\alpha}{16}\,
\left(7\left|\Phi_1\right|^2+
9\left|\Phi_2\right|^2
\right)
\frac{{\cal F}_B(x_2,\vec q_2)}
{(q_2^2+q_{2\,min}^2)^2}\ .
\label{10}
 \eeq
 Here $\vec p_T$ and $y$ are the transverse momentum and rapidity of the
produced quark (or antiquark); $x_q$ is the fraction of the plus component
of the momentum of the hadron $h_1$ taken by the quark, which is related
to the rapidity interval $\Delta y=\ln(1/x_q)$ between the hadron and the
quark; and $\vec q_{1,2}$ are the transverse momenta of the gluons
radiated by the hadrons $h_{1,2}$ with light-cone fractional momenta
$x_{1,2}$ respectively. While we integrate over $x_1$, the value of $x_2$
is defined by the kinematics,
 \beq
x_2=\frac{1}{x_1s}\left[
\frac{m_q^2+p_T^2}{\alpha} + 
\frac{m_q^2+(\vec p_T -\vec q_1-\vec q_2)^2}
{1-\alpha}\right]\ ,
\label{20}
 \eeq
 where $\alpha=x_q/x_1$. The function $\Phi_{1}$ in (\ref{10})  
corresponds to the sum of diagrams Fig.~\ref{one-gluon}a,b, while
$\Phi_{2}$ corresponds to the difference of the amplitudes
Fig.~\ref{one-gluon}a,b plus the graph in Fig.~\ref{one-gluon}c. These
functions are expressed in terms of the usual LC wave functions
$\Psi^G_{\bar qq}(\vec k_T,\alpha,q^2)$ of the $\bar qq$ Fock state in a
gluon, where $\vec k_T$ is the relative transverse momentum of the $\bar
qq$,
 \beq
\Psi^G_{\bar qq}(\vec k_T,\alpha,q_1^2) = 
\sqrt{\frac{4\alpha_s}{3}}\,
\frac{\chi^+_q\,\hat O\,\tilde\chi^*_{\bar q}}
{\epsilon^2+k_T^2}\ ,
\label{30}
 \eeq
where
\beqn
\epsilon^2 &=& \alpha(1-\alpha)Q_1^2+m_q^2\ ;\nonumber\\
Q_1^2 &=& q_1^2+q_{1\,min}^2\ \ \ \ 
(q_{min}\sim\Lambda_{QCD})\ ;\nonumber\\
\hat O &=& m\vec\sigma\cdot\vec e + (1-2\alpha)
(\vec\sigma\cdot\vec n)(\vec p\cdot e) +
i(\vec p\times\vec n)\cdot\vec e\ ;
\label{40}
 \eeqn
 and the gluon polarization vector is related to its transverse momentum,
$\vec e=\vec q_1/q_1$. Then, we have
 \beqn
\Phi_1 &=& \Psi^G_{\bar qq}(\vec p_T-\alpha\vec q_1,\alpha,q_1^2) -
\Psi^G_{\bar qq}(\vec p_T-\alpha\vec q_1-\vec q_2,\alpha,q_1^2)\ ;
\nonumber\\
\Phi_2 &=& \Psi^G_{\bar qq}(\vec p_T-\alpha\vec q_1,\alpha,q_1^2) +
\Psi^G_{\bar qq}(\vec p_T-\alpha\vec q_1-\vec q_2,\alpha,q_1^2)-
2\,\Psi^G_{\bar qq}(\vec p_T-\alpha\vec q_1-\alpha\vec q_2,\alpha,q_1^2).
\label{50}
 \eeqn

\subsection{Multiple interactions}\label{multi-gluon}

In the light-cone approach employed in the rest frame of the target (the
bottom hadron in Fig.~\ref{one-gluon}) the process depicted in
Fig.~\ref{one-gluon} looks like the interaction of a $\bar qq$ fluctuation
of the projectile gluon with the target. Although the interaction is
mediated by one gluon exchange, one can make it more realistic using the
phenomenological dipole cross section $\sigma^h_{\bar qq}(r,x)$ of
interaction of a $\bar qq$ dipole of transverse separation $\vec r$ with a
hadron $h$ at energy $s\sim (xr^2)^{-1}$. This cross section fitted to
data incorporates the unknown dynamics of soft multi-gluon exchanges and
radiation. Applying to (\ref{10}) a Fourier transformation we get,
 \beqn
&&\frac{d\sigma(A\,B\to q\,X)}{d^2p_T\,dy} =
\frac{1}{8\pi^2}
\int\limits_{x_q}^1 dx_1\,\frac{\alpha}{x_1}\,
\int d^2q_1\,d^2q_2\,
\frac{q_1^2\,{\cal F}_A(x_1,\vec q_1)}
{(q_1^2+q_{1\,min}^2)^2}
\nonumber\\ &\times& 
\int d^2r_1\,d^2r_2\,
\exp\Bigl[i(\vec p_T-\alpha\vec q_2)
(\vec r_1-\vec r_2)\Bigr]\,
\Psi^G_{\bar qq}\Bigr.^\dagger(\vec r_2,\alpha,q_1^2)\,
\Psi^G_{\bar qq}(\vec r_1,\alpha,q_1^2)\,
\Sigma^B(\vec r_1,\vec r_2,x_2,\alpha),
\label{60}
 \eeqn
 where
\beqn
\Sigma^B(\vec r_1,\vec r_2,x_2,\alpha) &=&
{1\over16}\,\left\{9\,\Bigl[
\sigma_{\bar qq}^B(\vec r_1-\alpha\vec r_2,x_2)+
\sigma_{\bar qq}^B(\vec r_2-\alpha\vec r_1,x_2)+
\sigma_{\bar qq}^B(\alpha\vec r_1,x_2)+
\sigma_{\bar qq}^B(\alpha\vec r_2,x_2)\Bigr] 
\right. \nonumber\\ &-& \left. 
\Bigl[\sigma_{\bar qq}^B(\vec r_1,x_2) +
\sigma_{\bar qq}^B(\vec r_2,x_2) -
8\sigma_{\bar qq}^B(\vec r_1-\vec r_2,x_2)-
8\sigma_{\bar qq}^B(\alpha\vec r_1-\alpha\vec r_2,x_2)
\Bigr]
\right\}.
\label{70}
 \eeqn
 At $\vec r_1=\vec r_2$ this becomes the familiar combination
$\sigma_3(\vec r,\alpha,x) = {9\over8}\{(\sigma^h_{\bar qq}(\alpha\vec
r,x) +\sigma^h_{\bar qq}[(1-\alpha)\vec r,x]\} - {1\over8}\sigma^h_{\bar
qq}[(2\alpha-1)\vec r,x]$ which is the dipole cross section for a
three-body system $\bar qqG\ra$ interacting with a hadron $h$. In
particular, it enters the total cross section of $\bar qq$ pair production
by a gluon \cite{npz,hir,kt-charm}.

The dipole cross section vanishes at small $\bar qq$ separations,
$\sigma^h_{\bar qq}(r,x)\Bigr._{r_T\to0} =
r^2\,G^h(x,1/r^2)\,\pi^2\alpha_s/3$, where $G^h(x,1/r^2) = xg^h(x,1/r^2)$
is the gluon distribution function in hadron $h$. In this limit the dipole
cross section corresponds to one gluon exchange (in the inelastic
amplitude) and Eq.~(\ref{10}) is recovered. At the same time, it is
usually assumed that at large separations $\sigma^h_{\bar qq}(r_T,x)$
saturates at some constant value $\sigma^h_0$. This may be motivated by
either saturation of the gluon density, or shortness of the gluon
interaction radius.

Let us introduce a function 
 \beq
\omega^h(\vec r,x)=\sigma^h_0-\sigma_{\bar qq}^h(\vec r,x)\ ,
\label{80}
 \eeq
 which has the properties $\omega^h(\vec r,x)_{r\to0}\to\sigma^h_0(x)$ and
$\omega(\vec r,x)_{r\to\infty}\to0$. Therefore its Fourier transform,
 \beq
\omega^h(\vec q,x)=\frac{1}{(2\pi)^2}\int
d^2r\,\omega^h(\vec r,x)\,e^{i\vec q\cdot\vec r}\ ,
\label{90}
 \eeq
 is defined for any $\vec q$. Then, taking into account that $\int
d^2q\,\omega^h(\vec q,x)\,e^{-i\vec q\cdot\vec r}=\sigma_0^h(x)$, we can
represent the function $\Sigma^B(\vec r_1,\vec r_2,\alpha)$,
Eq.~(\ref{70}), in the form,
 \beqn
\Sigma^B(\vec r_1,\vec r_2,x_2\alpha) &=&
\int d^2q\,\omega^B(q,x_2)\,\left\{{7\over16}\,
\left[1-e^{i\vec q\cdot\vec r_1}\right]
\left[1-e^{-i\vec q\cdot\vec r_2}\right]\right.
\nonumber\\ &+&
\left.{9\over16}\,
\left[1+e^{i\vec q\cdot\vec r_1}
-2\,e^{i\alpha\vec q\cdot\vec r_1}\right]
\left[1+e^{i\vec q\cdot\vec r_2}
-2\,e^{i\alpha\vec q\cdot\vec r_2}\right]
\right\}\ .
\label{100}
 \eeqn
 Using this expression we can rewrite Eq.~(\ref{60}) as
 \beq
\frac{d\sigma(A\,B\to q\,X)}{d^2p_T\,dy} =
\frac{1}{2}
\int\limits_{x_q}^1 dx_1\,\frac{\alpha}{x_1}\,
\int d^2q_1\,
\frac{q_1^2\,{\cal F}_A(x_1,\vec q_1)}
{(q_1^2+q_{1\,min}^2)^2}\ \omega^B(\vec q,x_2)\,
\left({7\over16}\left|\Phi_1\right|^2+
{9\over16}\left|\Phi_2\right|^2
\right)
\label{110}
 \eeq

Finally, using the relation $\omega(\vec q,x)=\sigma_0(x)\delta(\vec q) -
\sigma_{\bar qq}(\vec q,x)$ we arrive at,
 \beq
\frac{d\sigma(A\,B\to q\,X)}{d^2p_T\,dy} =
- \frac{1}{2}
\int\limits_{x_q}^1 dx_1\,\frac{\alpha}{x_1}\,
\int d^2q_1\,
\frac{q_1^2\,{\cal F}_A(x_1,\vec q_1)}
{(q_1^2+q_{1\,min}^2)^2}\,
\left({7\over16}\left|\Phi_1\right|^2+
{9\over19}\left|\Phi_2\right|^2\right)\,
\sigma^B_{\bar qq}(\vec q,x_2)\,
\label{115}
 \eeq
 Comparing this expression with Eq.~(\ref{10}) we conclude that
the transformations done above are equivalent to the replacement 
 \beq
\frac{q_2^2\,{\cal F}_B(x_2,\vec q_2)}
{(q_2^2+q_{2\,min}^2)^2}\ \Rightarrow\ 
-\,\frac{3}{4\pi\alpha_s}\,
\sigma_{\bar qq}^B(\vec q_2,x_2)
\label{120}
 \eeq
 in Eq.~(\ref{10}). This observation leads to the natural assumption that
the same procedure should be performed with the contribution to
Eq.~(\ref{10}) of the upper part of the graphs in Fig.~\ref{one-gluon},
namely,
 \beq
\frac{q_1^2\,{\cal F}_A(x_1,\vec q_1)}
{(q_1^2+q_{1\,min}^2)^2}\ \Rightarrow\ 
-\,\frac{3}{4\pi\alpha_s}\,
\sigma_{\bar qq}^A(\vec q_1,x_1)\ .
\label{130}
 \eeq

Switching back to coordinate representation,
 \beq
\sigma^A_{\bar qq}(\vec q,x) = 
\frac{1}{(2\pi)^2}\int d^2\rho\,
e^{-i\vec q\cdot\vec\rho}\,
\vec\nabla_\rho^2\sigma_{\bar qq}^A(\vec\rho,x)\ ,
\label{140}
 \eeq
 we eventually arrive at the cross section in a form which includes
multi-gluon exchange,
 \beqn
\frac{d\sigma(A\,B\to q\,X)}{d^2p_T\,dy} &=&
\frac{6}{(4\pi)^3\alpha_s}\,
\int\limits_{x_q}^1 dx_1\,\frac{\alpha}{x_1}
\int d^2q \int d^2\rho\,d^2r_1\,d^2r_2\,
\exp\Bigl[i(\vec p_T-\alpha\vec q)\cdot(\vec r_1-\vec r_2) -
i\vec q\cdot\vec\rho\Bigr]\nonumber\\
&\times& \vec\nabla_\rho^2\sigma^A_{\bar qq}(\vec\rho,x_1)\,
\Psi_{\bar qq}^G\Bigr.^\dagger(\vec r_2,\alpha,q^2)\,
\Psi_{\bar qq}^G\Bigr.(\vec r_1,\alpha,q^2)\,
\Sigma^B(\vec r_1,\vec r_2,x_2,\alpha)\ .
\label{150}
 \eeqn
 This is the central result of this paper. It looks asymmetric, while the
graphs in Fig.~\ref{one-gluon} are symmetric relative to beam-target
interchange. This expression, however, has been derived in the rest frame
of the target. In the beam rest frame one should just interchange
$A\Leftrightarrow B$ and $x_1\Leftrightarrow x_2$.

The total yield of quarks integrated over transverse momentum has
the simple form,
 \beq
\frac{d\sigma(A\,B\to q\,X)}{dy} =
\frac{6}{(4\pi)^3\alpha_s}\,
\int\limits_{x_q}^1 dx_1\,\frac{\alpha}{x_1}
\int d^2q \int d^2\rho\,d^2r\,
e^{-i\vec q\cdot\vec\rho}\,
\vec\nabla_\rho^2\sigma^A_{\bar qq}(\vec\rho,x_1)\,
\left|\Psi_{\bar qq}^G\Bigr.(\vec r,\alpha,q^2)\right|^2
\sigma_3^B(\vec r,\alpha)\ .
\label{160}
 \eeq

Thus the production cross section is expressed in terms of the dipole
cross section either on one (A), or another (B) colliding particles or
nuclei. Unfortunately, it has an asymmetric form which is related to our
choice of $\alpha_A$. Switching to $\alpha_B$ we will get an equivalent
cross section, but having a different form.

\section{Summary}

Guided by the abelian analogue of particle production in QED, we suggested
an approach incorporating multiple interactions both in the beam and
target, within the dipole formalism. The proposed procedure of replacing
the unintegrated gluon density in the target by a combination of cross
sections of dipole-target interaction is assumed to be valid also for the
beam. The main result, Eq.~(\ref{150}), still needs needs to be tested
through numerical calculations and comparison with data. This expression
looks asymmetric relative to the dipole cross sections of interacting with
the colliding particles or nuclei, $A$ and $B$. However, effectively it is
symmetric provided that a proper replacement of variables is done. It is
possible to rewrite it in an explicitly symmetric form, however, in this
case the light-cone distribution function $\Psi_{\bar qq}$ in (\ref{150})
should be replaced by a more complicated function. Further development of
this formalism will be published elsewhere.

\bigskip

\noindent
 {\bf Acknowledgment}: We are grateful to Hans-J\"urgen Pirner and Andreas
Sch\"afer for helpful and inspiring discussions. A.T. and O.V. thank the
Physics Departments of USM, Valparaiso, and Heidelberg University for
hospitality. This work has been partially supported by Fondecyt (Chile)
grant numbers 1050519, 1030355 and 7050175.


\begin{thebibliography}{99}

\bibitem{zkl} B.Z.~Kopeliovich, L.I.~Lapidus and A.B.~Zamolodchikov, Sov.
Phys. JETP Lett. {\bf 33} (1981) 595; Pisma v Zh. Exper. Teor.  Fiz.
{\bf 33} (1981) 612.

\bibitem{al} A.H. Mueller and B.~Patel, Nucl. Phys. {\bf B425} (1994) 471.


\bibitem{krt1} B.Z.~Kopeliovich, J.~Raufeisen and A.V.~Tarasov,
Phys. Lett. {\bf B440} (1998) 151.

\bibitem{krt2}  B.Z.~Kopeliovich, J.~Raufeisen and A.V.~Tarasov,
Phys. Rev. {\bf C62} (2000) 035204.

\bibitem{kps1} B.Z.~Kopeliovich, I.K.~Potashnikova and Ivan Schmidt, Phys.
Rev. {\bf C73} (2006) 034901.

\bibitem{bjorken} J.M.~Bjorken, J.B.~Kogut and D.E.~Soper, {\bf D3} (1971)
1382.

\bibitem{nz} N.N.~Nikolaev and B.G.~Zakharov,
Z. Phys. {\bf C49} (1991) 607

\bibitem{hir} B.Z. Kopeliovich {\sl Soft Component of Hard Reactions and
Nuclear Shadowing (DIS, Drell-Yan reaction, heavy quark production)}, in
proc. of the Workshop 'Dynamical Properties of Hadrons in Nuclear Matter',
Hirschegg 1995, ed. H. Feldmeier and W. Noerenberg, p. 102
(hep-ph/9609385).

\bibitem{km} Yu.V.~Kovchegov and A.H.~Mueller, Nucl.  Phys. B {\bf 529},
451 (1998).

\bibitem{kst1} B.Z.~Kopeliovich, A.~Sch\"afer and A.V.~Tarasov,
Phys. Rev. C {\bf 59} (1999) 1609.   

\bibitem{yuri} Yu.V.~Kovchegov, Nucl. Phys. {\bf A692} (2001) 557.

\bibitem{weiz} C.F. von Weizsacker, Z. Phys. {\bf 88} (1934) 612.

\bibitem{will} E.J. Williams, Phys. Rev. {\bf 45} (1934) 729.

\bibitem{npz} N.N. Nikolaev, G.~Piller, B.G. Zakharov, JETP {\bf 81} (1995)
851 [Zh. Eksp. Teor. Fiz. {\bf 108} (1995) 1554.

\bibitem{kt-charm} B.Z.~Kopeliovich and A.V.~Tarasov, Nucl. Phys. {\bf 
A710} (2002) 180.


\end{thebibliography}
\end{document}